\documentclass[conference]{IEEEtran}
\IEEEoverridecommandlockouts
\usepackage{amsmath,amssymb,amsfonts}
\usepackage{acronym}
\usepackage{algorithmic}
\usepackage{float}
\usepackage{graphicx}
\usepackage{textcomp}
\usepackage{xcolor}
\usepackage{url}
\usepackage{csquotes}
\def\BibTeX{{\rm B\kern-.05em{\sc i\kern-.025em b}\kern-.08em
    T\kern-.1667em\lower.7ex\hbox{E}\kern-.125emX}}
\usepackage[numbers]{natbib}
\usepackage{tikz}
\usetikzlibrary{decorations.text}
\usetikzlibrary{shapes}
\usepackage{pgfplots}
\usepackage{pgfplotstable}
\usepgfplotslibrary{groupplots}
\usepgfplotslibrary{fillbetween}
\usepackage{subfig}
\pgfplotsset{compat=1.16}

\begin{document}

\acrodef{EMCCD}[EMCCD]{electron-multiplying charge-coupled device}
\acrodef{CMOS}[CMOS]{complementary metal-oxide-semiconductor}
\acrodef{qCMOS}[qCMOS]{quantitative \ac{CMOS}}
\acrodef{MPQ}[MPQ]{Max Planck Institute for Quantum Optics}
\acrodef{sCIC}[sCIC]{serial clock-induced charge}
\acrodef{CIC}[CIC]{clock-induced charge}
\acrodef{SNR}[SNR]{signal-to-noise ratio}
\acrodef{EM}[EM]{electron-multiplying}
\acrodef{MTF}[MTF]{modulation transfer function}
\acrodef{PSF}[PSF]{point-spread function}
\acrodef{OTF}[OTF]{optical transfer function}
\acrodef{CDF}[CDF]{cumulative distribution function}
\acrodef{PDF}[PDF]{probability density function}
\acrodef{GCC}[GCC]{GNU Compiler Collection}
\acrodef{ROI}[ROI]{region of interest}
\acrodef{ADC}[ADC]{analog-to-digital converter}
\acrodef{MQV}[MQV]{Munich Quantum Valley}
\acrodef{PSF}[PSF]{point spread function}
\acrodef{RL}[RL]{Richardson-Lucy}
\acrodef{GNS}[GNS]{Gauss-Newton solver}
\acrodef{MLE}[MLE]{maximum likelihood estimator}
\acrodef{IO}[I/O]{input/output}

\title{Comparison of Atom Detection Algorithms for Neutral Atom Quantum Computing\thanks{This work was funded by the German Federal Ministry of Education and Research (BMBF) under the funding program \textit{Quantum Technologies - From Basic Research to Market} under contract numbers 13N16070, 13N16076, and 13N16087, as well as from the Munich Quantum Valley~(MQV), which is supported by the Bavarian State Government with funds from the Hightech Agenda Bayern.}
}

\author{\IEEEauthorblockN{Jonas Winklmann}
\IEEEauthorblockA{\textit{Chair of Computer Architecture and Parallel Systems}\\
\textit{Technical University of Munich}\\
Munich, Germany\\
jonas.winklmann@tum.de}
\and
\IEEEauthorblockN{Andrea Alberti}
\IEEEauthorblockA{\textit{Quantum Many-Body Systems Division}\\
\textit{Max Planck Institute of Quantum Optics}\\
Garching, Germany\\
andrea.alberti@mpq.mpg.de}
\and
\IEEEauthorblockN{Martin Schulz}
\IEEEauthorblockA{\textit{Chair of Computer Architecture and Parallel Systems}\\
\textit{Technical University of Munich}\\
Munich, Germany\\
schulzm@in.tum.de}
}

\bibliographystyle{IEEEtranN}

\IEEEpubid{\vspace{1cm}\begin{minipage}{\textwidth}\ \\[24pt] \centering
  979-8-3315-4137-8/24/\$31.00 \copyright 2024 IEEE. Personal use of this material is permitted. Permission from IEEE must be obtained for all other uses, in any current or future media, including reprinting/republishing this material for advertising or promotional purposes, creating new collective works, for resale or redistribution to servers or lists, or reuse of any copyrighted component of this work in other works. DOI 10.1109/QCE60285.2024.00124
\end{minipage}}

\IEEEaftertitletext{\vspace{-0.4\baselineskip}}

\maketitle

\begin{abstract}
In neutral atom quantum computers, readout and preparation of the atomic qubits are usually based on fluorescence imaging and subsequent analysis of the acquired image. For each atom site, the brightness or some comparable metric is estimated and used to predict the presence or absence of an atom. Across different setups, we can see a vast number of different approaches used to analyze these images. Often, the choice of detection algorithm is either not mentioned at all or it is not justified. 

We investigate several different algorithms and compare their performance in terms of both precision and execution run time. To do so, we rely on a set of synthetic images across different simulated exposure times with known occupancy states, which we generated using a previously validated imaging simulation. Since the use of simulation provides us with the ground truth of atom site occupancy, we can easily state precise error rates and variances of the reconstructed property.

However, knowing the relative performance of these algorithms is not sufficient to justify their use, since better ones can exist that were not compared. To investigate this possibility, we calculated the Cramér-Rao bound in order to establish an upper limit that even a perfect estimator cannot outperform. As the metric of choice, we used the number of photonelectrons that can be contributed to a specific atom site. Every estimator that reconstructs a different property can simply be scaled accordingly. Since the bound depends on the occupancy of neighboring sites, we provide the best and worst cases, as well as a half filled one, which should represent an averaged bound best.

Our comparison shows that of our tested algorithms, a global non-linear least-squares solver that uses the optical system's \ac{PSF} to return a global bias and each sites' number of photoelectrons performed the best, on average crossing the worst-case bound for longer exposure times. Its main drawback is its huge computational complexity and, thus, required calculation time. We manage to somewhat reduce this problem, suggesting that its use may be viable, leading us to a novel group of algorithms that present a compromise between speed and precision. However, our study also shows that for cases where utmost speed is required, simple algorithms, like summing up pixel values around atom sites, may be preferable.
\end{abstract}


\IEEEpubidadjcol
\section{Introduction}
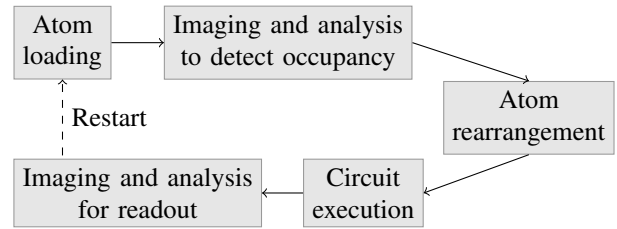
\begin{figure}
\centering
    \begin{tikzpicture}
        %Nodes
        \node (loading) at (0, 1) [draw=gray!80,fill=gray!20,align=center] {Atom\\loading};
        \node (imaging) at (3, 1) [draw=gray!80,fill=gray!20,align=center] {Imaging and analysis\\to detect occupancy};
        \node (sorting) at (6.2, 0) [draw=gray!80,fill=gray!20,align=center] {Atom\\rearrangement};
        \node (execution) at (4, -1) [draw=gray!80,fill=gray!20,align=center] {Circuit\\execution};
        \node (readout) at (1, -1) [draw=gray!80,fill=gray!20,align=center] {Imaging and analysis\\for readout};
        %Lines
        \draw[->] (loading.east) to (imaging.west);
        \draw[->] (imaging.east) to (sorting.north);
        \draw[->] (sorting.south) to (execution.east);
        \draw[->] (execution.west) to (readout.east);
        \draw[->, dashed] (0, -0.5) to node [right] {Restart} (loading.south);
    \end{tikzpicture}
    \caption{Simplified steps in compute cycle of neutral atom quantum computer}
    \label{figures:computeCycle}
\end{figure}
\begin{figure}
\centering
\includegraphics[width=0.7\linewidth]{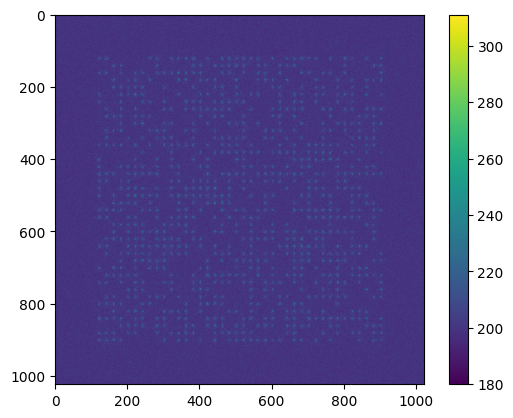}
\caption{Simulated exemplary image that reproduces \mbox{qCMOS} characteristics}
\label{figures:exampleImage}
\end{figure}
Neutral atom quantum computer typically rely on fluorescence imaging to detect atoms and to read qubit states~\cite{Morgado:1}. During preparation, atoms are loaded into a 2D grid, with each site effectively containing either zero or one atom, each with a given probability around 50\%. While the atoms are fluorescing, an image is taken with a highly accurate camera and analyzed in order to determine for each site whether it is occupied by an atom or not.

In fig.~\ref{figures:computeCycle}, we can see a simplified illustration of the steps involved in operating the quantum computer. Since imaging and subsequent image analysis has to occur twice during every compute cycle, it is vital to limit the required time to a minimum. As we can only continue with rearranging the atoms once at least a partial result of this image analysis is available, both the exposure time of the image, as well as the time taken to analyze the image, must be optimized. To be precise, we want to minimize the sum of these processes, since not all systems require similar exposure times.

For a fast readout that does not care about atom loss and that makes use of high photon scattering rates, a fast detection method may be better suited than one that extracts all possible information at the cost of a longer run time. In fact, readout on a timescale of only a few microseconds has recently been demonstrated~\cite{Scholl:1,Su:1}. In that case, acquiring more information during imaging may require less time than extracting that same information through advanced calculations. One can imagine that it is impractical to employ a detection algorithm that runs perhaps a thousand times slower than the imaging process. Lossless and gentle imaging, on the other hand, could minimize exposure time and atom heating caused by fluorescence imaging by using more complex atom detection algorithms.

At the moment, many systems, especially ones with large atom spacings, use a very simple method of summing up the pixels in an area around the atom site and determining its observed occupancy through thresholding~\cite{Kwon:1,Norcia:1,Cheuk:1}. While this is sufficient for most cases and even the best approach for some, it does not always minimize the total time spent on atom detection.

In this work, we, therefore, compare different detection algorithms in order to study the involved tradeoffs, both in terms of precision in atom occupancy detection and execution time of the detection algorithm. Additionally, we determine an absolute limit for the performance of a perfect algorithm through the Cramér-Rao bound and evaluate how close the investigated algorithms come to this limit. To determine the precision of the algorithms, we use a previously developed and validated simulator that generates realistic images starting with the ground truth occupancy~\cite{Winklmann:1}. Fig.~\ref{figures:exampleImage} shows an example of such an image. This allows us to easily state both the algorithms' error rates and the variances of their estimated metrics. Further, we investigate the temporal complexity and optimization potential of each algorithm.

In order to get an overview of the plethora of popular algorithms, we follow this section by exploring other publications and grouping their approaches into categories. Next, we establish the Cramér-Rao bound, so we can subsequently compare the algorithms against this limit in section~\ref{sec:detection}. Afterwards, we compare their error rates and run times and conclude by recommending the use of different approaches depending on the situation.

Our contributions include:
\begin{itemize}
    \item We investigate and collect the most prevalent atom detection approaches.
    \item We establish a limit for the precision of a perfect estimator.
    \item We benchmark our selected algorithms against each other and against the previously established bound.
    \item We analyse the tradeoff between speed and accuracy and provide advise on how to choose the best-suited method.
\end{itemize}

Our tests show that the simple, yet most widely used, algorithm that only sums up all pixels around an atom site is comparatively inaccurate. However, the low conceptual and computational complexity still makes it applicable for use cases where run time is important or where optimization of more complex algorithms is impossible. Overall, a non-linear least-squares solver proved to be the most precise method tested. For higher photoelectron counts, it even comes close to satisfying the limits of a perfect estimator. Although its main downside of requiring a substantial run time may prevent it from being used in real situations in its current state, we found several optimization opportunities and hope to be able to reduce its calculation time further in the future. Until this step is achieved, algorithms such as the state-reconstruction library, \ac{RL}, or Wiener deconvolution may present a compelling compromise between speed and precision and can, depending on the use case, lower overall time spent on atom detection. 
\section{Related Work}\label{chapters:relatedWork}
Many research groups require accurate single-atom detection from fluorescence imaging. However, few have compared different methods or justified their choice of algorithm. One notable exception is the work of~\citeauthor{Rooij:1}, who have also gone the route of using simulated images to compare different atom detection methods, albeit for much smaller site spacings than we work with. This work aims at going further in establishing an absolute bound for the precision of these methods, as well as using more different algorithms. They compare \ac{RL} and Wiener deconvolution, as well as a local-iterative method, with the former two clearly outperforming the latter~\cite{Rooij:1}. As such, \ac{RL} and Wiener deconvolution are also discussed in our work. Since we are dealing with image analysis, it seems natural to investigate the use of these well-established techniques in order to reduce the influence of the optical system and concentrate each atom sites' emissions onto as few pixels as possible. Unsurprisingly, deconvoluting the image using the \ac{PSF} of the optical system is also popular in other neutral-atom-related projects~\cite{Miranda:1, Yamamoto:1}.

To get some perspective on which atom detection methods are used in other works, several different prevalent approaches are mentioned in the following.

The most obvious method is to sum up the values in some region of interest around each atom site. While this approach is very simple, it has an extremely low algorithmic complexity, allowing it to be used in cases where any difference in calculation time is vital. It seems to be commercially implemented in Quantum Machines' control unit for neutral atoms, although their wording of a \enquote{straightforward region of interest image processing procedure} leaves some room to speculation~\cite{qm:1}. Also, it is a popular method used in many research projects~\cite{Kwon:1,Norcia:1,Cheuk:1}. We refer to this method in the following as the \ac{ROI} method. \citeauthor{Madjarov:1} introduces a slight variation of this method in his thesis, mentioning a weighted sum with weights taken from the \ac{PSF} \cite{Madjarov:1}.

Another category of detection algorithms are fitting algorithms. By fitting some local or global modeling function to the image data, one can acquire a set of function parameters, such as amplitudes or emission counts, which can be used to determine occupancy, usually via thresholding. This is also a method that is popular in many research projects~\cite{Parsons:1, Alberti:1}. In our work, we investigate this method by finding each site's emission parameter as the result of a global least-squares problem, which we solve by using a weighted Gauss-Newton solver.

In our work on developing the image simulation tool, we mentioned how having a way of acquiring labelled data enables one to use machine leaning to predict occupancy \cite{Winklmann:1}. While our own research into this topic has not yet been fruitful, an autoencoder-based method has recently been shown to be promising. In their approach, \citeauthor{Impertro:1} show how one can use deep learning to eliminate the need for labelled data. In the encoder stage, the data is compressed until each atom site is represented by only one value at the bottleneck layer. The network is trained by using this compressed data to reproduce the initial image as closely as possible in the decoder stage \cite{Impertro:1}. 

\Acp{MLE} are another interesting idea, which might be worth investigating further, since there have been publications showing an improvement over summation and fitting methods \cite{Martinez-Dorantes:1, Burrell:1}. Incidentally, our work on Cramér-Rao bounds already provides us with some insights into the likelihood function and its Hessian matrix. However, this is not yet covered within the scope of this project.
\section{Cramér–Rao Bound}
Any relative comparison between detection algorithms leaves the question of whether better performance can somehow be achievable. To give a bound on the performance of a perfect estimator, we rely on the Cramér–Rao bound. This gives us a goal, since no algorithm can reliably achieve a better precision. We consider an estimator that uses all available information to predict the number of photoelectrons caused by emissions from each individual atom site within the exposure time. If an estimator maps the information onto a different property, we can just scale it accordingly in order to be comparable.

\subsection{Deriving the Cramér–Rao Bound}
\label{chapters:cramerRaoCompute}
As we can pretty much eliminate column, row, and pixel noise via calibration, we ignore these noise sources here. Let us, therefore, assume a Poissonian distributed photoelectron distribution that takes into account background illumination and dark current. We divide the resulting number of electrons by the preamp gain, add a global pixel offset, and use the result as the mean for the readout distribution. The overall probability distribution of a pixel value is the sum over the scaled readout distributions for all possible photoelectron counts. Thus, we define the resulting probability $p(q,x,\gamma)$ for pixel $x$ to assume value $q$ as
\begin{equation}
    p(q,x,\gamma) = \frac{1}{\sigma\sqrt{2\pi}}\cdot P*G(q,x,\gamma)
\end{equation}
with the convolution of the Poissonian distribution $P(x,\gamma,k)$ and the non-normalized Gaussian distribution $G(q,k)$ described by
\begin{equation}
    P*G(q,x,\gamma) = \sum_{k=0}^\infty P(x,\gamma,k)\cdot G(q,k)
\end{equation}
\begin{equation}
    P(x,\gamma,k) = \frac{\lambda(x,\gamma)^k\cdot e^{-\lambda(x,\gamma)}}{k!}
\end{equation}
\begin{equation}
    \lambda(x,\gamma) = b + \sum_{i=0}^nPSF_i(x) \cdot\gamma_i
    \label{equation:pixelValue}
\end{equation}
\begin{equation}
    G(q,k) = e^{-\frac{(q-(o+\frac{k}{g}))^2}{2\sigma^2}}
\end{equation}
with $\gamma\in\mathbb{R}^n$ denoting the total number of photoelectrons that are expected to be received from each of the $n$ atom sites. $PSF_i(x)$ represents the fraction of photoelectrons caused by atom site $i$ that end up at pixel $x$. The sum of each \ac{PSF} over all pixels should be $1$. $\lambda(x,\gamma)$, therefore, represents the expected number of photoelectrons at a given pixel, which we use as the parameter to the Poissonian distribution. We then use the global pixel offset $o$ and preamp gain $g$ to convert the electron count to a digital signal, while the readout standard deviation $\sigma$ describes the noise involved in the readout.

The Cramér-Rao bound depends on the Fisher information matrix of the observation of an atom. Each pixel value can be seen as an independent measurement of an atom site's brightness, even if the site is far from the pixel. The total information can be acquired by summing over each measurement's individual Fisher information matrix. Therefore, we define it as the following sum over all $m$ pixels:
\begin{equation}
    [I(\gamma)]_{i,j} = \sum_{x=0}^{m}\int_\mathbb{R}\frac{\frac{\partial}{\partial\gamma_i}p(q,x,\gamma)\cdot\frac{\partial}{\partial\gamma_j}p(q,x,\gamma)}{p(q,x,\gamma)}dq
\end{equation}
with the partial derivative of the pixel function being described by
\begin{equation}
\begin{aligned}
    \frac{\partial}{\partial\gamma_i}p(q,x,\gamma) = \frac{1}{\sigma\sqrt{2\pi}}\cdot\frac{\partial}{\partial\gamma_i}P*G(q,x,\gamma)\\
    =\frac{1}{\sigma\sqrt{2\pi}}\cdot\sum_{k=0}^\infty(P(x,\gamma,k)\cdot G(q,k)\cdot f_i(x,\gamma,k))
\end{aligned}
\end{equation}
We multiply each summand of $P*G(q,x,\gamma)$ with the factor
\begin{equation}
    f_i(x,\gamma,k) = \frac{PSF_i(x)\cdot (k - \lambda(x,\gamma))}{\lambda(x,\gamma)}
\end{equation}
to get the corresponding derivative.
The variance of an estimator T is then bound by
\begin{equation}
    var(T_i) \ge [I(\gamma)^{-1}]_{i,i}
\end{equation}

\subsection{Computing the Cramér–Rao Bound}

\begin{figure}
\centering
  \subfloat[Empty (NN)]{\includegraphics[width=0.33\linewidth]{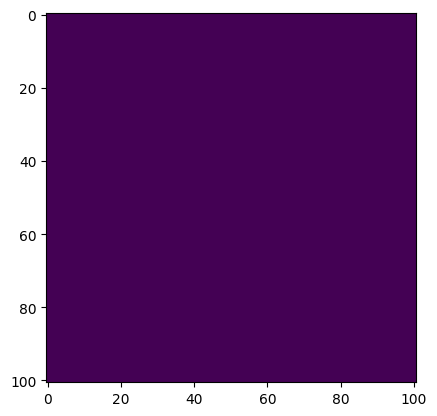}}
  \subfloat[Empty (SN)]{\includegraphics[width=0.33\linewidth]{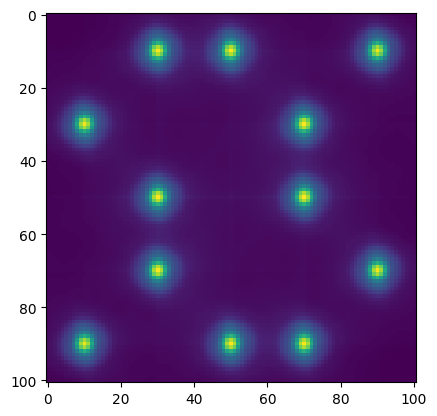}}
  \subfloat[Empty (AN)]{\includegraphics[width=0.33\linewidth]{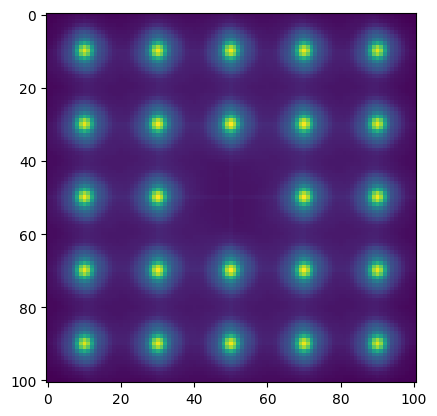}}\\
  \subfloat[Occupied (NN)]{\includegraphics[width=0.33\linewidth]{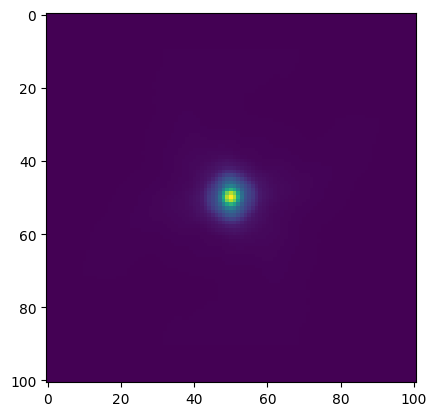}}
  \subfloat[Occupied (SN)]{\includegraphics[width=0.33\linewidth]{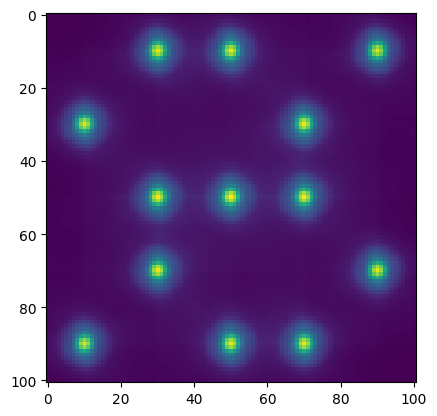}}
  \subfloat[Occupied (AN)]{\includegraphics[width=0.33\linewidth]{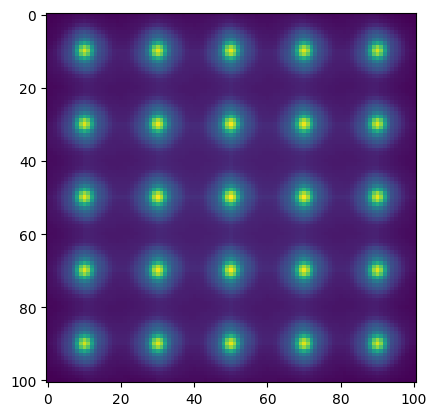}}
\caption{Test scenarios for atom configurations considered for the Cramér-Rao bound: (a) Empty central site, no surrounding atoms; (b) Empty central site, some surrounding atoms; (c) Empty central site, all surrounding sites occupied; (d) Occupied central site, no surrounding atoms (this essentially shows the \ac{PSF} with an empty border around it); (e) Occupied central site, some surrounding atoms; and (f) Occupied central site, all surrounding sites occupied}\label{figures:cramerRaoCases}
\end{figure}

As a compromise between calculation time and precision, we chose a \ac{PSF} of size 81x81 pixels. Any values outside this area are negligible and so is any benefit that an estimator may draw from using such distant pixel values. In total, we use an image with 101x101 pixels, containing 25 atom sites in a 5x5 configuration, spaced apart 20 pixels. As, in practice, most atoms are not at the edge of the image, we only investigate the central atom site. Any atoms that are at the edge will actually be easier to detect, somewhat loosening the bound for smaller atom arrays.

Since the amount of different occupancy configurations makes it practically impossible to calculate every possible one, we consider six different cases. The atom site may be empty or occupied and there may be no neighbors, a half-filled configuration, or a fully-occupied neighborhood. Fig.~\ref{figures:cramerRaoCases} shows these different scenarios. 

If there are many images with random occupancy, the overall expected distribution of reconstructed states is the weighted sum of the distributions for each given neighborhood. For high occupancy rates, we would, therefore, expect the actual limit to be closer to the fully-occupied-neighborhood case and vice versa. The half-filled configuration should give us a good estimate on the achievable limit for arbitrary configurations with 50\% fill rate.

For two atom sites $i$ and $j$, the corresponding entry $[I(\gamma)]_{i,j}$ in the Fisher information matrix is the sum of the Fisher information for each pixel that is in range of both $i$'s and $j$'s \ac{PSF}. Since analytical solutions to the problem are not available, we need to estimate this information. To calculate a pixel's probability distribution, we compute $p(q,x,\gamma)$ for all values $k$ for which the Poissonian contribution $P(x,\gamma,k)$ is greater than an $\epsilon=10^{-8}$. Any more precise values did not notably change any results in testing and the thresholds were readily available through the Boost library's Poissonian quantile and inverse survival function~\cite{Boost:1}, leading to reasonable execution times. In order to integrate over all possible pixel values, we calculate the result for each $q$, for which the Gaussian distribution $G(q,k)$ is greater than $\epsilon=10^{-8}$, in intervals of $0.1$ and then interpolate them linearly. Thereby, the readout Gaussian, which is centered on the average expected pixel value $o+\frac{k}{g}$, is cut off symmetrically at both sides.

By choosing to have the relevant pixel values $q$ depend on the value of $k$, we cannot simply calculate $\frac{\frac{\partial}{\partial\gamma_i}p(q,x,\gamma)\cdot\frac{\partial}{\partial\gamma_j}p(q,x,\gamma)}{p(q,x,\gamma)}$ for a given $q$, but have to keep track of the results for all $q$ while iterating over $k$. However, since each value of $k$ is treated equally due to the range of considered values for $q$ being centered on the average expected outcome, this leads to more accurate results than having a range of $q$ that is independent of $k$.

\pgfplotsset{height=0.46\textwidth,width=\axisdefaultwidth,compat=1.9}
\begin{figure}
\centering
\begin{tikzpicture}
\begin{axis}[
    xlabel={Photoelectrons per atom site},
    ylabel=Minimum variance,
    ymin=0,xmin=0,
    legend pos=north west,
    legend entries={Empty (NN),Empty (SN),Empty (AN),Occupied (NN),Occupied (SN),Occupied (AN)},
    ]
  \addplot[blue,mark=*] table [x=photons,y=emptyE] {data/varianceBound.dat};
  \addplot[red,mark=*] table [red,x=photons,y=fiftyE] {data/varianceBound.dat};
  \addplot[green,mark=*] table [green,x=photons,y=fullE] {data/varianceBound.dat};
  \addplot[blue,dashed,mark=square,mark options={solid}] table [blue,dotted,x=photons,y=emptyOcc] {data/varianceBound.dat};
  \addplot[red,dashed,mark=square,mark options={solid}] table [x=photons,y=fiftyOcc] {data/varianceBound.dat};
  \addplot[green,dashed,mark=square,mark options={solid}] table [x=photons,y=fullOcc] {data/varianceBound.dat};
\end{axis}
\end{tikzpicture}
\caption{Minimally achievable variance of estimator for empty and filled sites. NN = No Neighbors, SN = Some Neighbors (50\% filled exemplary usecase), AN = All Neighbors\label{figures:cramerRaoBound}}
\end{figure}
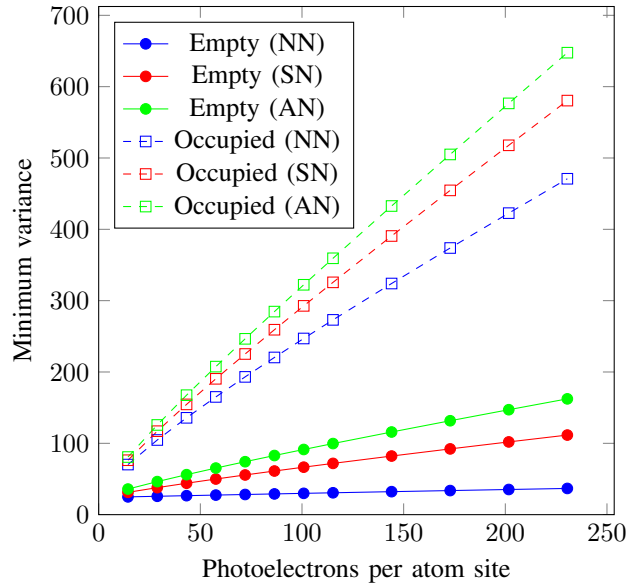
\subsection{Results}
Fig.~\ref{figures:cramerRaoBound} shows the minimally achievable variance of an estimator that reconstructs the number of photoelectrons caused by each atom site. As the x-axis, we chose the number of photoelectrons contributed to an atom site. In order to precisely compare the precision of detection algorithms with the bound, there is a data point for each of the reconstructed photoelectron counts from the algorithm test data set. We can see that a site containing an atom has a much higher variance than an empty one, which makes sense due to the higher number of photons and, thus, the larger parameter for the Poissonian shot noise.

\pgfplotsset{height=0.45\textwidth,width=\axisdefaultwidth,compat=1.9}
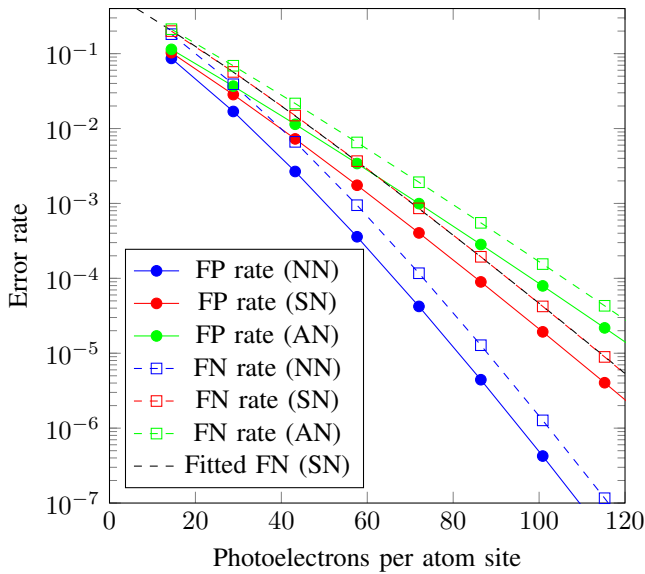
\begin{figure}
\centering
\begin{tikzpicture}
\begin{axis}[
    xlabel={Photoelectrons per atom site},
    ylabel=Error rate,
    legend pos=south west,
    xmin=0,xmax=120,ymin=0.0000001,ymax=0.4,
    legend entries={FP rate (NN),FP rate (SN),FP rate (AN),FN rate (NN),FN rate (SN),FN rate (AN), Fitted FN (SN)},
    ymode=log,
    declare function={erf(\x)=%
      (1+(e^(-(\x*\x))*(-265.057+abs(\x)*(-135.065+abs(\x)%
      *(-59.646+(-6.84727-0.777889*abs(\x))*abs(\x)))))%
      /(3.05259+abs(\x))^5)*(\x>0?1:-1);}
    ]
  \addplot[blue,mark=*] table [x=photons,y=emptyFp] {data/varianceBound.dat};
  \addplot[red,mark=*] table [red,x=photons,y=fiftyFp] {data/varianceBound.dat};
  \addplot[green,mark=*] table [green,x=photons,y=fullFp] {data/varianceBound.dat};
  \addplot[blue,dashed,mark=square,mark options={solid}] table [blue,dotted,x=photons,y=emptyFn] {data/varianceBound.dat};
  \addplot[red,dashed,mark=square,mark options={solid}] table [x=photons,y=fiftyFn] {data/varianceBound.dat};
  \addplot[green,dashed,mark=square,mark options={solid}] table [x=photons,y=fullFn] {data/varianceBound.dat};
  \addplot[black,dashed] plot[domain=0:120,samples=200] {0.5*(1+erf((0.391*x^0.951+2.160-x)/sqrt(2*(4.183*x^0.896+31.618))};
\end{axis}
\end{tikzpicture}
\caption{Minimally achievable error rates when setting the threshold so that overall error is minimized assuming 50\% fill rate. FP = false positive, FN = false negative, NN = No Neighbors, SN = Some Neighbors (50\% filled exemplary use case), AN = All Neighbors\label{figures:cramerRaoBoundError}}
\end{figure}

Fig.~\ref{figures:cramerRaoBoundError} shows the error rates that a perfect estimator can produce. In this case, the number of empty and occupied sites are the same and the threshold is set at the intersection point of the distributions for empty and occupied ones, thus minimizing the total error rate. It can be seen that the false negative rate is always higher than the false positive rate. This makes intuitive sense since the variance of the distribution for empty sites is much narrower. As a result, it is more likely for an occupied site to appear dim than for an empty site to appear bright.

The relationship between error rate and photoelectron count can be tremendously useful when designing a quantum computer setup, since there may be requirements to reach given maximal error rates, e.g., for some error correction scheme to be viable. Knowing how many photoelectrons are required to reach this goal allows one to design the setup accordingly.

Let try, as an example, to approximate the false negative rate of the case where some atoms are present. Firstly, we require functions describing the behaviour of threshold and variance with increasing photoelectron counts. We use $a\cdot k^b + c$ as the fitting function with $k$ denoting the number of photoelectrons and $a$, $b$, and $c$ three fitting parameters. In our case, the threshold can be estimated quite precisely by {$0.391\cdot k^{0.951}+2.160$} and the achievable variance for this case follows \mbox{$4.183\cdot k^{0.896}+31.618$} closely. If we insert these functions into a Gaussian's \ac{CDF}, we can estimate the false negative rate
\begin{equation}
    \text{P(\enquote{FN})}\approx\frac{1+erf(\frac{0.391\cdot x^{0.951}+2.160}{\sqrt{2\cdot(4.183\cdot x^{0.896}+31.618)}})}{2}
\end{equation}
The black dashed line in fig.~\ref{figures:cramerRaoBoundError} shows this function. We can see that the real bound perfectly matches our expection.
\section{Detection Methods}
\label{sec:detection}
In order to compare the tested methods, we use simulated images generated by a validated simulator capable of constructing realistic images from a given ground truth. Each image has a size of 1024x1024 pixels and contains 40x40 atom sites (i.e, 1600 sites) with a spacing of 20 pixels between sites. Configured with a scattering rate of 28\,000 photons per second per atom, a numerical aperture of $\text{NA}=0.65$, and a quantum efficiency of $0.86$, we would expect up to $28\,000\cdot 0.86\cdot \frac{1-\sqrt{1-\text{NA}^2}}{2}\approx2890$ photonelectrons per second. There are 12 datasets of 100 images each, with simulated exposure times of 5ms, 10ms, 15ms, 20ms, 25ms, 30ms, 35ms, 40ms, 50ms, 60ms, 70ms, and 80ms.

There will be graphs showing the algorithms' variances. In each of these figures, the six graphs from Fig.~\ref{figures:cramerRaoBound}, which represent the Cramér-Rao bound, will be shown as two gray bands, one for empty and one for occupied sites, with a black centerline representing the case where some neighboring atoms are present.
\subsection{Region of Interest}
\label{chapters:roi}
The first and most simple detection method is the \ac{ROI} method, in which we take the sum over all pixels within a region around each atom site and threshold to determine occupancy. Here we use all pixels with a distance of less or equal to 5 pixels.
\pgfplotsset{height=0.33\textwidth,width=\axisdefaultwidth,compat=1.9}
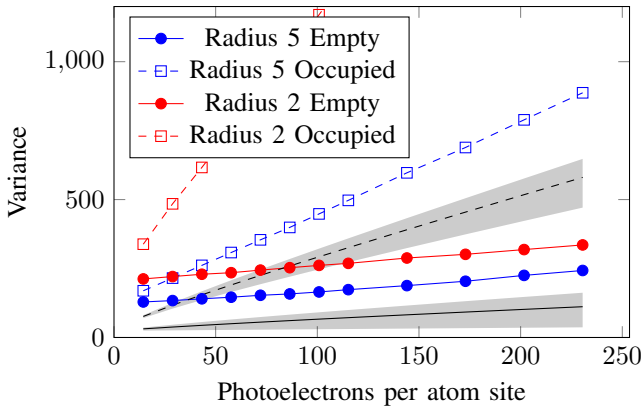
\begin{figure}
\centering
\begin{tikzpicture}
\begin{axis}[
    xlabel={Photoelectrons per atom site},
    ylabel=Variance,
    legend pos=north west,ymax=1200,ymin=0,xmin=0,
    legend entries={Radius 5 Empty, Radius 5 Occupied, Radius 2 Empty, Radius 2 Occupied}
    ]
  \addplot[blue,mark=*] table [x=photons,y=roi5] {data/varByAlgorithmEmpty.dat};
  \addplot[blue,dashed,mark=square,mark options={solid}] table [x=photons,y=roi5] {data/varByAlgorithmOccupied.dat};
  \addplot[red,mark=*] table [x=photons,y=roi2] {data/varByAlgorithmEmpty.dat};
  \addplot[red,dashed,mark=square,mark options={solid}] table [x=photons,y=roi2] {data/varByAlgorithmOccupied.dat};
  \addplot[name path=lowerE, draw=none] table [x=photons,y=emptyE] {data/varianceBound.dat};
  \addplot[black] table [red,x=photons,y=fiftyE] {data/varianceBound.dat};
  \addplot[name path=upperE, draw=none] table [green,x=photons,y=fullE] {data/varianceBound.dat};
  \addplot[name path=lower, draw=none] table [blue,dotted,x=photons,y=emptyOcc] {data/varianceBound.dat};
  \addplot[black,dashed] table [x=photons,y=fiftyOcc] {data/varianceBound.dat};
  \addplot[name path=upper, draw=none] table [x=photons,y=fullOcc] {data/varianceBound.dat};
  \addplot[fill=black, forget plot, fill opacity=0.2] fill between[of=lower and upper];
  \addplot[fill=black, forget plot, fill opacity=0.2] fill between[of=lowerE and upperE];
\end{axis}
\end{tikzpicture}
\caption{Variances of reconstruction results of test images using the \ac{ROI} method\label{figures:resultsROI}}
\end{figure}
Fig.~\ref{figures:resultsROI} shows the variances that the \ac{ROI} method was able to achieve. It is apparent that the method does not come close to the established limits, but it also does not perform terrible if the radius of the \ac{ROI} is chosen correctly. Further, this method can only even be considered if the spacing between atoms allows for larger radii.

Its run time is directly proportional to the number of atom sites and the number of pixels within the radius. The implementation used here simply takes the sum of the corresponding values sequentially in Python. Naturally, this is not very optimal. A fast C or C++ method could be a lot faster and a GPU or FPGA implementation should be purely \ac{IO} bound.
\subsection{Wiener}\label{chapters:wiener}
The Wiener filter aims at reconstructing an unknown underlying signal from noisy data \cite{Wiener:1}. One of its most prevalent use cases lies in image deconvolution, which is also what we are using it for. For simplicity, we used the \textit{unsupervised\_wiener} and \textit{wiener} functions from the scikit-image library. The latter of the two has a tunable \textit{balance} parameter \cite{Skimage:1}. For this, we tried several different values. In the following, we simply append this value to the algorithm name, e.g. \textit{Wiener 20} represents the \textit{wiener} function with balance parameter 20.
\pgfplotsset{height=0.6\textwidth,width=\axisdefaultwidth,compat=1.9}
\begin{figure}
\centering
\begin{tikzpicture}
\begin{axis}[
    xlabel={Photoelectrons per atom site},
    ylabel=Variance,
    legend pos=north west,ymin=0,xmin=0,ymax=1100,
    legend entries={Uns. Wiener Empty, Uns. Wiener Occupied,Wiener 35 Empty, Wiener 35 Occupied,Wiener 10 Empty, Wiener 10 Occupied,Wiener 3 Empty, Wiener 3 Occupied}
    ]
  \addplot[blue,mark=*] table [x=photons,y=wiener] {data/varByAlgorithmEmpty.dat};
  \addplot[blue,dashed,mark=square,mark options={solid}] table [x=photons,y=wiener] {data/varByAlgorithmOccupied.dat};
  \addplot[red,mark=*] table [x=photons,y=wiener35] {data/varByAlgorithmEmpty.dat};
  \addplot[red,dashed,mark=square,mark options={solid}] table [x=photons,y=wiener35] {data/varByAlgorithmOccupied.dat};
  \addplot[brown,mark=*] table [x=photons,y=wiener10] {data/varByAlgorithmEmpty.dat};
  \addplot[brown,dashed,mark=square,mark options={solid}] table [x=photons,y=wiener10] {data/varByAlgorithmOccupied.dat};
  \addplot[green,mark=*] table [x=photons,y=wiener3] {data/varByAlgorithmEmpty.dat};
  \addplot[green,dashed,mark=square,mark options={solid}] table [x=photons,y=wiener3] {data/varByAlgorithmOccupied.dat};
  \addplot[name path=lowerE, draw=none] table [x=photons,y=emptyE] {data/varianceBound.dat};
  \addplot[black] table [red,x=photons,y=fiftyE] {data/varianceBound.dat};
  \addplot[name path=upperE, draw=none] table [green,x=photons,y=fullE] {data/varianceBound.dat};
  \addplot[name path=lower, draw=none] table [blue,dotted,x=photons,y=emptyOcc] {data/varianceBound.dat};
  \addplot[black,dashed] table [x=photons,y=fiftyOcc] {data/varianceBound.dat};
  \addplot[name path=upper, draw=none] table [x=photons,y=fullOcc] {data/varianceBound.dat};
  \addplot[fill=black, forget plot, fill opacity=0.2] fill between[of=lower and upper];
  \addplot[fill=black, forget plot, fill opacity=0.2] fill between[of=lowerE and upperE];
\end{axis}
\end{tikzpicture}
\caption{Variances of reconstruction results of test images using the Wiener deconvolution\label{figures:resultsWiener}}
\end{figure}
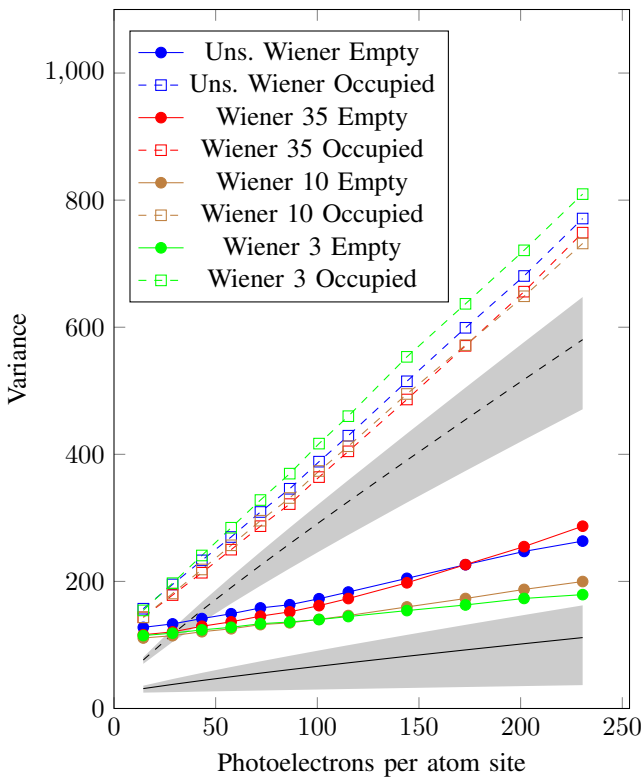
Fig.~\ref{figures:resultsWiener} shows the variances that the Wiener deconvolution was able to achieve. It is apparent that no one value for the \textit{balance} parameter is perfect, since higher values can deconvolve occupied sites better and lower values are better suited for empty sites. The best value, therefore, depends on the occupancy rate. For our case of around 50\%, we found that a parameter of around 10 produced the lowest error rates.

Since the performance of the Wiener deconvolution is limited by the performance of the required Fourier transform, we can assume its run-time class to be $\mathcal{O}(m\cdot log(m))$ for an image containing $m$ pixels.
\subsection{Richardson-Lucy}\label{chapters:rl}
\ac{RL} deconvolution is another popular method to reconstruct an image that has been convoluted with a known \ac{PSF}. It is an iterative approach and is based on Bayesian statistics \cite{Richardson:1, Lucy:1}. Again, we did not develop the algorithm ourselves but used the implementation from the scikit-image library \cite{Skimage:1}. The only parameter that requires tuning is the number of iterations and, again, we will append the number of iterations to the abbrevation \textit{\ac{RL}}.
\pgfplotsset{height=0.44\textwidth,width=\axisdefaultwidth,compat=1.9}
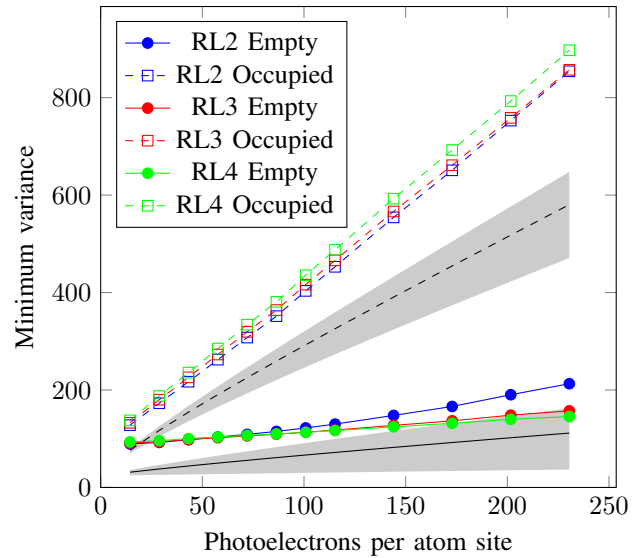
\begin{figure}
\centering
\begin{tikzpicture}
\begin{axis}[
    xlabel={Photoelectrons per atom site},
    ylabel=Minimum variance,
    legend pos=north west,ymin=0,xmin=0,
    legend entries={RL2 Empty, RL2 Occupied,RL3 Empty, RL3 Occupied,RL4 Empty, RL4 Occupied}
    ]
  \addplot[blue,mark=*] table [x=photons,y=rl2] {data/varByAlgorithmEmpty.dat};
  \addplot[blue,dashed,mark=square,mark options={solid}] table [x=photons,y=rl2] {data/varByAlgorithmOccupied.dat};
  \addplot[red,mark=*] table [x=photons,y=rl3] {data/varByAlgorithmEmpty.dat};
  \addplot[red,dashed,mark=square,mark options={solid}] table [x=photons,y=rl3] {data/varByAlgorithmOccupied.dat};
  \addplot[green,mark=*] table [x=photons,y=rl4] {data/varByAlgorithmEmpty.dat};
  \addplot[green,dashed,mark=square,mark options={solid}] table [x=photons,y=rl4] {data/varByAlgorithmOccupied.dat};
  \addplot[name path=lowerE, draw=none] table [x=photons,y=emptyE] {data/varianceBound.dat};
  \addplot[black] table [red,x=photons,y=fiftyE] {data/varianceBound.dat};
  \addplot[name path=upperE, draw=none] table [green,x=photons,y=fullE] {data/varianceBound.dat};
  \addplot[name path=lower, draw=none] table [blue,dotted,x=photons,y=emptyOcc] {data/varianceBound.dat};
  \addplot[black,dashed] table [x=photons,y=fiftyOcc] {data/varianceBound.dat};
  \addplot[name path=upper, draw=none] table [x=photons,y=fullOcc] {data/varianceBound.dat};
  \addplot[fill=black, forget plot, fill opacity=0.2] fill between[of=lower and upper];
  \addplot[fill=black, forget plot, fill opacity=0.2] fill between[of=lowerE and upperE];
\end{axis}
\end{tikzpicture}
\caption{Variances of reconstruction results of test images using the \ac{RL} deconvolution\label{figures:resultsRL}}
\end{figure}

Usually, one wants to choose a sufficiently high iteration count for the image to converge to the unblurred one. However, fig.~\ref{figures:resultsRL} shows that this is not necessarily the best approach here. We can see that less iterations are actually preferred for occupied atom sites, while more iterations perform better for empty ones.

In each iteration, the algorithm has to iterate over every pixel of the image and do a multiplication for every other pixel within the its \ac{PSF}'s range. Therefore, its run time scales linearly with the number of iterations, the number of pixels in an image, and the number of pixels in the \ac{PSF}.
\subsection{State Reconstruction}
\label{chapters:sr}
Further, we investigate~\citeauthor{Wei:1}'s state-reconstruction library~\cite{Wei:1}. It is to be noted that this algorithm is mainly aimed at smaller lattice spacings. Since these lattices are prone to drifting, it also incorporates a phase reconstruction that tries to find the global shift of less than half an atom site, for which the emission is maximized. For other algorithms, this detection is omitted. Instead, they use the known atom site locations. Since a detection apparatus may be tuned and calibrated beforehand, we deem it acceptable to assume the atom locations to be known precisely.
\pgfplotsset{height=0.35\textwidth,width=\axisdefaultwidth,compat=1.9}
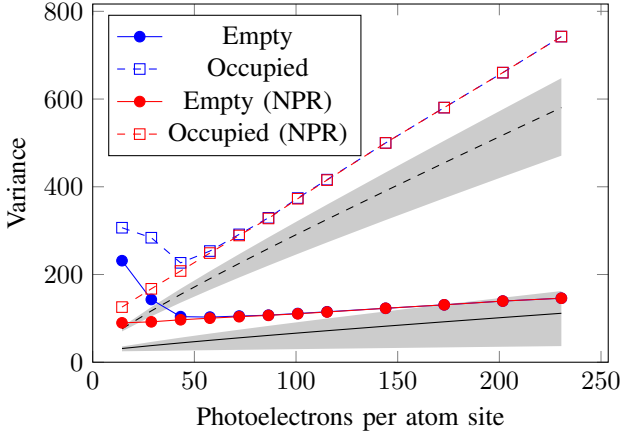
\begin{figure}
\centering
\begin{tikzpicture}
\begin{axis}[
    xlabel={Photoelectrons per atom site},
    ylabel=Variance,
    legend pos=north west,ymin=0,xmin=0,
    legend entries={Empty, Occupied, Empty (NPR), Occupied (NPR)}
    ]
  \addplot[blue,mark=*] table [x=photons,y=er] {data/varByAlgorithmEmpty.dat};
  \addplot[blue,dashed,mark=square,mark options={solid}] table [x=photons,y=er] {data/varByAlgorithmOccupied.dat};
  \addplot[red,mark=*] table [x=photons,y=erNP] {data/varByAlgorithmEmpty.dat};
  \addplot[red,dashed,mark=square,mark options={solid}] table [x=photons,y=erNP] {data/varByAlgorithmOccupied.dat};
  \addplot[name path=lowerE, draw=none] table [x=photons,y=emptyE] {data/varianceBound.dat};
  \addplot[black] table [red,x=photons,y=fiftyE] {data/varianceBound.dat};
  \addplot[name path=upperE, draw=none] table [green,x=photons,y=fullE] {data/varianceBound.dat};
  \addplot[name path=lower, draw=none] table [blue,dotted,x=photons,y=emptyOcc] {data/varianceBound.dat};
  \addplot[black,dashed] table [x=photons,y=fiftyOcc] {data/varianceBound.dat};
  \addplot[name path=upper, draw=none] table [x=photons,y=fullOcc] {data/varianceBound.dat};
  \addplot[fill=black, forget plot, fill opacity=0.2] fill between[of=lower and upper];
  \addplot[fill=black, forget plot, fill opacity=0.2] fill between[of=lowerE and upperE];
\end{axis}
\end{tikzpicture}
\caption{Variances of reconstruction results of test images using the state-reconstruction library, NPR = No phase reconstruction\label{figures:resultsSR}}
\end{figure}

Fig.~\ref{figures:resultsSR} shows that this library performs quite well, especially for cases with longer exposure times. We can see that the two graphs labelled \textit{Empty} and \textit{Occupied} have higher variances for very low photoelectron counts, only matching up with their counterparts for higher ones. These blue graphs show the algorithm's default behaviour. However, after noticing inconsistencies in the phase detection and realizing that it is not required, we disabled it, which lead to the two red graphs. Now, the algorithm consistently outperforms both \ac{RL} and Wiener deconvolution.
\subsection{Gauss-Newton Solver}\label{chapters:gns}
Since all simulated images rely on the same \ac{PSF}, the \ac{PSF} does not need to be estimated by a non-linear solver. Instead, we can imagine a linear solver that globally solves occupation by adding the given \acp{PSF} to a bias. However, each pixel is subject to different levels of noise, depending on the observed brightness. If these weights are to be considered, then the solver does become non-linear. We use a Gauss-Newton solver that takes the emission rate at each site, as well as the global pixel offset value $o$ as parameters. We define the parameter vector $\beta\in\mathbb{R}^{n+1}$ as 
\begin{equation}
    \beta_i= 
\begin{cases}
    \gamma_i,       & \text{if } i\leq n\\
    o,              & \text{if } i=n+1
\end{cases}
\end{equation}
Each pixel value is modeled as in (\ref{equation:pixelValue}). The derivative of the image modeling function with respect to an atom site $i$ is simply the \ac{PSF} of $i$. We can define the modeling function's Jacobian matrix $J$ to be the $m\times (n+1)$ matrix with elements
\begin{equation}
    j_{ij}= 
\begin{cases}
    PSF_i(j),       & \text{if } i\leq n\\
    1,              & \text{if } i=n+1
\end{cases}
\end{equation}
The last column is a bit bothersome to deal with, but handles the global bias $o$ that is added to each pixel. One could also add additional columns for row and column noise if that were to be considered. We can see that this approach is also suited for images where each atom site has its own \ac{PSF}, as the Jacobian can take this into account. Doing so will not impact performance, since this can be done in advance.

According to~\citeauthor{Yang:1}, the iteration step can be defined as
\begin{equation}
    \beta^{(s+1)} = \beta^{(s)} - (J^TWJ)^{-1}J^TWr(\beta^{(s)})
\end{equation}
where $W\in\mathbb{R}^m$ represents the vector of weights for each pixel, with the weights being the inverse of each data points variance \cite{Yang:1}. Since the readout and shot noise are uncorrelated, the variance of the value of pixel $x$ is given by $Var(x)=\sigma^2 + c(x,\gamma)$. The residual function $r(\beta^{(s)})$ is defined as the difference between the original image and the output of the model function using current parameter estimations.
\pgfplotsset{height=0.28\textwidth,width=\axisdefaultwidth,compat=1.9}
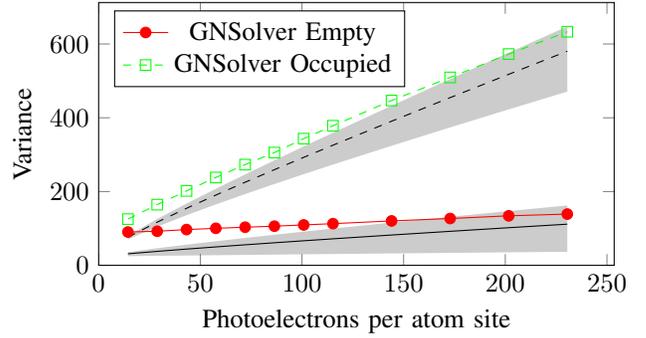
\begin{figure}
\centering
\begin{tikzpicture}
\begin{axis}[
    xlabel={Photoelectrons per atom site},
    ylabel=Variance,
    legend pos=north west,ymin=0,xmin=0,
    legend entries={GNSolver Empty, GNSolver Occupied}
    ]
  \addplot[red,mark=*] table [x=photons,y=gns] {data/varByAlgorithmEmpty.dat};
  \addplot[green,dashed,mark=square,mark options={solid}] table [x=photons,y=gns] {data/varByAlgorithmOccupied.dat};
  \addplot[name path=lowerE, draw=none] table [x=photons,y=emptyE] {data/varianceBound.dat};
  \addplot[black] table [red,x=photons,y=fiftyE] {data/varianceBound.dat};
  \addplot[name path=upperE, draw=none] table [green,x=photons,y=fullE] {data/varianceBound.dat};
  \addplot[name path=lower, draw=none] table [blue,dotted,x=photons,y=emptyOcc] {data/varianceBound.dat};
  \addplot[black,dashed] table [x=photons,y=fiftyOcc] {data/varianceBound.dat};
  \addplot[name path=upper, draw=none] table [x=photons,y=fullOcc] {data/varianceBound.dat};
  \addplot[fill=black, forget plot, fill opacity=0.2] fill between[of=lower and upper];
  \addplot[fill=black, forget plot, fill opacity=0.2] fill between[of=lowerE and upperE];
\end{axis}
\end{tikzpicture}
\caption{Variances of reconstruction results of test images using a Gauss-Newton solver\label{figures:resultsGNS}}
\end{figure}

The drawback of this algorithm is its computational complexity. On our test machine, the average calculation time for the initial version was about 37.75s. However, we found some optimization possibilities that might lead to this approach being viable after all. Firstly, the image can be split into subregions, which decreases the size of both dimensions of the Jacobian matrix. Furthermore, the weights and, therefore, also the product of $J^TWJ$, converges along with the parameters. Calculating these only every few iterations decreased calculation times further to an average of 4.15s. Reducing the \ac{PSF} dimensions to 41x41 leads to an average of 1.84s while influencing precision only marginally. Apart from optimizing the algorithm itself, one could also imagine a simple way of estimating the results and using this guess as the starting condition. This could reduce the number of required iterations notably. While this approach is still too slow to be usable, it indicates that this algorithm might be viable for some applications once optimized to run on GPUs or FPGAs, assuming the necessary computational capabilities are present. 

In practice, the run time of the Gauss-Newton solver is limited by constructing $J^TWJ$ and finding its inverse, which is done using the Conjugate Gradient method. The former part has to iterate over all atom sites with overlapping \acp{PSF}, resulting in a run time of $\mathcal{O}(n\cdot q^2)$, since the number of relevant atom sites scales with the number of pixels in the \ac{PSF}. Finding its inverse can be done in $\mathcal{O}(n^\frac{3}{2})$ \cite{Shewchuk:1}.

This method also reliably provides the actual photoelectron count per atom. For each data set, we fit a Gaussian distribution to the distributions of the reconstructed property for both empty and occupied sites. Taking the difference and multiplying with the preamp gain results in the amount of photoelectrons per atom. Since this value should be directly proportional to the exposure time, a straight line was again fitted to the relation between exposure time and photoelectrons, resulting in a rate of around 2881 counts per second. Accounting for a small loss due to aberrations, this value perfectly lines up with the previously established maximum of 2890. This rate of 2881 was used for establishing the Cramér-Rao bound and to scale the algorithms' variances appropriately.
\section{Comparison and Discussion}
\label{chapters:evaluation}
\subsection{Precision}
Having established both the theoretically optimal performance of detection algorithms, as well as the variances of all tested ones, we can now investigate their relative performance and how close they come to the bound.
\pgfplotsset{height=0.37\textwidth,width=\axisdefaultwidth,compat=1.9}
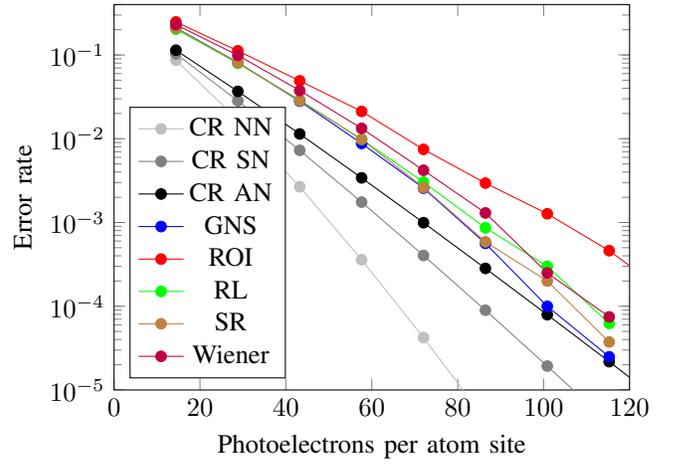
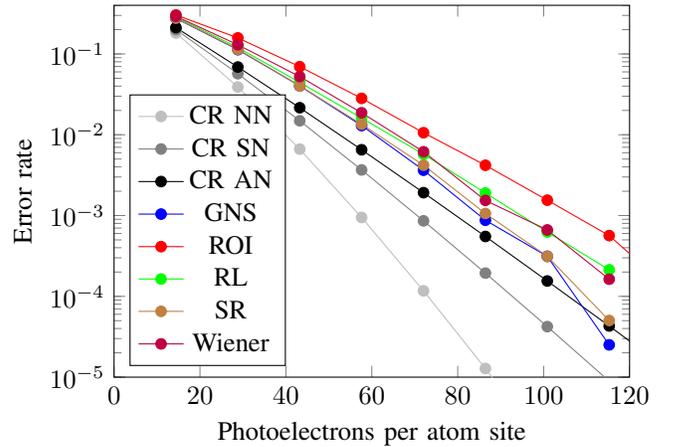
\begin{figure}
\centering
\subfloat[False positive rates]{
\begin{tikzpicture}
\begin{axis}[
    xlabel={Photoelectrons per atom site},
    ylabel=Error rate,
    legend pos=south west,
    xmin=0,xmax=120,ymin=0.00001,ymax=0.4,
    legend entries={CR NN,CR SN,CR AN,GNS,ROI,RL,SR,Wiener},
    ymode=log
    ]
  \addplot[gray!50,mark=*] table [x=photons,y=emptyFp] {data/varianceBound.dat};
  \addplot[gray,mark=*] table [red,x=photons,y=fiftyFp] {data/varianceBound.dat};
  \addplot[black,mark=*] table [red,x=photons,y=fullFp] {data/varianceBound.dat};
  \addplot[blue,mark=*] table [x=photons,y=gns] {data/falsePositiveRateByAlg.dat};
  \addplot[red,mark=*] table [x=photons,y=roi5] {data/falsePositiveRateByAlg.dat};
  \addplot[green,mark=*] table [x=photons,y=rl2] {data/falsePositiveRateByAlg.dat};
  \addplot[brown,mark=*] table [x=photons,y=erNP] {data/falsePositiveRateByAlg.dat};
  \addplot[purple,mark=*] table [x=photons,y=wiener10] {data/falsePositiveRateByAlg.dat};
\end{axis}
\end{tikzpicture}}\\
\pgfplotsset{height=0.36\textwidth,width=\axisdefaultwidth,compat=1.9}
\subfloat[False negative rates]{
\begin{tikzpicture}
\begin{axis}[
    xlabel={Photoelectrons per atom site},
    ylabel=Error rate,
    legend pos=south west,
    xmin=0,xmax=120,ymin=0.00001,ymax=0.4,
    legend entries={CR NN,CR SN,CR AN,GNS,ROI,RL,SR,Wiener},
    ymode=log
    ]
  \addplot[gray!50,mark=*] table [blue,dotted,x=photons,y=emptyFn] {data/varianceBound.dat};
  \addplot[gray,mark=*] table [x=photons,y=fiftyFn] {data/varianceBound.dat};
  \addplot[black,mark=*] table [x=photons,y=fullFn] {data/varianceBound.dat};
  \addplot[blue,mark=*] table [x=photons,y=gns] {data/falseNegativeRateByAlg.dat};
  \addplot[red,mark=*] table [x=photons,y=roi5] {data/falseNegativeRateByAlg.dat};
  \addplot[green,mark=*] table [x=photons,y=rl2] {data/falseNegativeRateByAlg.dat};
  \addplot[brown,mark=*] table [x=photons,y=erNP] {data/falseNegativeRateByAlg.dat};
  \addplot[purple,mark=*] table [x=photons,y=wiener10] {data/falseNegativeRateByAlg.dat};
\end{axis}
\end{tikzpicture}}
\caption{a) False positive and b) false negative rates for different algorithms, as well as the error rates given by the Cramér-Rao bound (CR (NN = No Neighbors, SN = Some Neighbors, AN = All Neighbors)), GNS = Gauss-Newton solver, ROI = region of interest, RL = Richardson-Lucy, SR = State Reconstruction, FP = false positive, FN = false negative\label{figures:errorRates}}
\end{figure}
Fig.~\ref{figures:errorRates} shows the false positive and false negative rates of the best-performing version of each of the tested algorithms, along with the error rates derived from the different Cramér-Rao bounds. We can see that the non-linear least-squares solver performs the best, closely followed by~\citeauthor{Wei:1}'s state-reconstruction. The Wiener and \ac{RL} deconvolution perform quite similar, clearly outperforming the \ac{ROI} method, especially for higher photoelectron counts.

The precision of these graphs is limited by the number of atom sites per dataset. As previously stated in Sec.~\ref{sec:detection}, there are 100 images with 1600 atom sites each, with approximately half being occupied and half empty. The error rates, therefore, have to be multiples of around $1.25\cdot 10^{-5}$.
\subsection{Run time}
During the detailed explanations of the individual algorithms, we also stated an expected run-time class wherever available. Unfortunately, the complexity of the unsupervised Wiener deconvolution and the state-reconstruction method did not allow such a specification.
\begin{table}[H]
    \centering
    \caption{Run time of detection algorithms by number of pixels per image $m$ and number of pixels in the \ac{PSF} $q$, number of atom sites $n$}
    \label{tab:runtimes}
    \begin{tabular}{ccc}
        \hline
        Algorithm & Average run time & Run-time class\\
        \hline
        Richardson-Lucy 2 iterations & 144.364ms & $\mathcal{O}(m\cdot q\cdot\#iter)$\\
        Richardson-Lucy 3 iterations & 212.750ms & $\mathcal{O}(m\cdot q\cdot\#iter)$\\
        Richardson-Lucy 4 iterations & 283.506ms & $\mathcal{O}(m\cdot q\cdot\#iter)$\\
        Richardson-Lucy 5 iterations & 358.175ms & $\mathcal{O}(m\cdot q\cdot\#iter)$\\
        Richardson-Lucy 6 iterations & 466.173ms & $\mathcal{O}(m\cdot q\cdot\#iter)$\\
        Gauss-Newton solver & 4146.611ms & $\mathcal{O}(n^\frac{3}{2}+n\cdot q^2)$\\
        Region of interest & 11.801ms & $\mathcal{O}(n)$\\
        state-reconstruction & 63.377ms & \\
        Wiener unsupervised & 1326.531ms & \\
        Wiener 3 & 42.327ms & $\mathcal{O}(m\cdot log(m))$\\
        Wiener 5 & 42.250ms & $\mathcal{O}(m\cdot log(m))$\\
        Wiener 10 & 42.224ms & $\mathcal{O}(m\cdot log(m))$\\
        Wiener 15 & 42.203ms & $\mathcal{O}(m\cdot log(m))$\\
        Wiener 35 & 42.189ms & $\mathcal{O}(m\cdot log(m))$\\
        Wiener 60 & 42.191ms & $\mathcal{O}(m\cdot log(m))$\\
        \hline
    \end{tabular}
\end{table}
Table~\ref{tab:runtimes} shows the run times and run-time classes for the tested methods as they have been derived in section~\ref{sec:detection}.

We can see that for cases with many atom sites, the \ac{RL} and Wiener deconvolution method can be useful, since their complexity does not depend on the number of sites. However, this raises the question of whether one could find a variation of these algorithms that does not treat every pixel as a potential atom location. Still, both algorithms are conceptually not very complex and can be well optimized.

At the cost of a very long execution time, one can consider using the Gauss-Newton solver. However, the state-reconstruction library offers a very similar performance while being much faster. However, further optimizing this method will not be trivial.
\section{Conclusions}
In this project, we compared different atom detection algorithms for neutral atom quantum computers and used the Cramér-Rao bound to establish a limit that even a perfect estimator cannot outperform. 

We saw that the global non-linear least-squares solver produced the best result at the cost of a long run time. The state-reconstruction library came in second, while also taking much less time. Its main drawback is its conceptual complexity, making it far more complicated to further optimize than other algorithms. 

Next are the Wiener and \ac{RL} deconvolution. Both of these deal with the image itself, independent of the atom sites. That means that they scale well with a higher number of atom sites. The precision and run time behavior of the two deconvolution methods was quite similar. \ac{RL} took slightly longer to run but also produced slightly better results.

The \ac{ROI} method performed the worst, but is also both computationally as well as conceptually by far the simplest method. Its use definitely can be justified for cases where speed is of utmost importance. One should, however, carefully check whether this approach fits the use case in order to not waste precision.

In general, we also saw that none of the methods come close to satisfying the Cramér-Rao bound for low photoelectron counts. While the \ac{GNS} crossed the worst-case bound at around 175 photoelectrons, it was far from the bound for lower counts. All other algorithms performed even worse, only sometimes also approaching the bound at higher counts. Unfortunately, we are more interested in the area around perhaps 60 to 100 photoelectrons, where the bound would already allow for error rates as low as $10^{-3}$ to $10^{-5}$. None of the tested algorithms are quite able to deliver such precision.

In the future, we also want to look into some other algorithms, such as variational autoencoders or other Bayesian methods. In terms of employing the methods, we will investigate the possibility of further improving the runtime requirements of the best performing ones. Once that is done, we will be able to make a decision based on precision, runtime, and imaging parameters of the target system.

\section*{Acknowledgments}
We thank the members of the Strontium Rydberg Lab at \ac{MPQ} and \ac{MPQ}'s \ac{MQV} team for many inspiring discussions on fluorescence imaging and atom detection.

\bibliography{bibliography.bib}

\end{document}